\documentclass[a4paper,pre,reqno,superscriptaddress,twocolumn, floatfix]{revtex4}
\usepackage{graphicx,color}
\usepackage{dcolumn}
\usepackage{epsfig}  
\usepackage[centertags]{amsmath}
\usepackage{amsfonts}
\usepackage{euscript}
\usepackage{amssymb}
\usepackage{amsthm}
\usepackage{newlfont}
\usepackage{lipsum}
\usepackage{mathtools}
\usepackage{mathrsfs}
\usepackage{subfigure}
\usepackage{todonotes}
\usepackage{kbordermatrix}
\usepackage[normalem]{ulem}
\usepackage[colorlinks=true,linkcolor=blue,urlcolor=blue,citecolor=blue,bookmarks=true]{hyperref}
\usepackage[utf8]{inputenc} 
\usepackage{comment}
\usepackage{float}
\begin{document}
 
\title{Evolutionary game selection creates cooperative environments}

\author{Onkar Sadekar}
\affiliation{Department of Network and Data Science, Central European University Vienna, Vienna 1100, Austria}

\author{Andrea Civilini}
\affiliation{School of Mathematical Sciences, Queen Mary University of London, London E1 4NS, United Kingdom}
\affiliation{Dipartimento di Fisica ed Astronomia, Universit\`a di Catania and INFN, Catania I-95123, Italy}

\author{Jesús Gómez-Gardeñes}
\affiliation{Department of Condensed Matter Physics, University of Zaragoza, 50009 Zaragoza, Spain}
\affiliation{GOTHAM lab, Institute of Biocomputation and Physics of
Complex Systems (BIFI), University of Zaragoza, 50018 Zaragoza, Spain}
\affiliation{Center for Computational Social Science, University of Kobe, 657-8501 Kobe, Japan}

\author{Vito Latora}
\affiliation{School of Mathematical Sciences, Queen Mary University of London, London E1 4NS, United Kingdom}
\affiliation{Dipartimento di Fisica ed Astronomia, Universit\`a di Catania and INFN, Catania I-95123, Italy}
\affiliation{Complexity Science Hub Vienna, A-1080 Vienna, Austria}

\author{Federico Battiston}
\affiliation{Department of Network and Data Science, Central European University Vienna, Vienna 1100, Austria}

\begin{abstract}
The emergence of collective cooperation in competitive environments is a well-known phenomenon in biology, economics, and social systems. While most evolutionary game models focus on the evolution of strategies for a fixed game, how strategic decisions co-evolve with the environment has so far mostly been overlooked. Here, we consider a game selection model where not only the strategies but also the game can change over time following evolutionary principles. Our results show that co-evolutionary dynamics of games and strategies can induce novel collective phenomena, fostering the emergence of cooperative environments. When the model is taken on structured populations the architecture of the interaction network can significantly amplify pro-social behaviour, with a critical role played by network heterogeneity and the presence of clustered groups of similar players, distinctive features observed in real-world populations. By unveiling the link between the evolution of strategies and games for different structured populations, our model sheds new light on the origin of social dilemmas ubiquitously observed in real-world social systems.
\end{abstract}

\maketitle

\section{Introduction}

Evolutionary Game Theory (EGT) combines game theory with Darwinian principles of natural selection and has made substantial contributions to diverse fields such as behavioural economics, social science, and biology \cite{smith_logic_1973, axelrod_evolution_1981, maynard_smith_evolution_1982, nowak_evolutionary_1992, nowak_supercooperators_2012}. Traditionally, EGT focuses on a population of players involved in a fixed game, evolving strategies over time, based on individual fitness \cite{hofbauer_evolutionary_1998, taylor_evolutionary_1978,  nowak_five_2006, szabo_evolutionary_2007}. Fitness is generally considered a growing function of the player's payoff, and it depends on the game and strategies of interacting players. This approach has yielded valuable insights into social and biological interactions, particularly in the context of social dilemmas—a widely recognized framework for investigating cooperation. In a social dilemma, each member of a group faces a choice: to cooperate or defect. Cooperation benefits the group but comes at an individual cost, while defectors enjoy the collective benefits without incurring any personal sacrifice. Classic examples of social dilemmas include the Prisoner's Dilemma and the Snowdrift Game \cite{axelrod_evolution_1981, nowak_evolutionary_1992}.

When applied to understanding human behaviour and puzzling aspects of social interactions, the simplified well-mixed approach of classical evolutionary game theory often leads to unrealistic predictions, and therefore, complex networks are used to describe patterns of interactions among players  \cite{hofbauer_evolutionary_1998, taylor_evolutionary_1978,  nowak_five_2006, szabo_evolutionary_2007, richter_half_2023}. This approach has proven instrumental in examining dynamical processes on real social networks, where players interact with their neighbours in a network structure \cite{santos_scale-free_2005,santos_evolutionary_2006,gomez-gardenes_dynamical_2007,perc_evolutionary_2013}. The concept of ``centrality" of a player in a network, quantifying its importance or influence, has been explored to understand the spread of information, the diffusion of innovations, and the formation of social norms \cite{perc_statistical_2017}. Moreover, traditional EGT models commonly rely on static, globally-defined payoff matrices \cite{wang_different_2014, glynatsi_bibliometric_2021}. This simplification fails to account for the reality that individual players engage in multiple social contexts and scenarios—each contributing to their overall fitness and subject to change over time.

In the last few years, these ideas have begun to emerge in the literature of evolutionary game theory. For instance, models based on stochastic games have been proposed to describe situations where players are in heterogeneous or periodically switching game environments \cite{solan_stochastic_2015, hilbe_evolution_2018, su_evolutionary_2019, amaral_heterogeneity_2020, galeazzi_ecological_2021, shu_eco-evolutionary_2022, roy_eco-evolutionary_2022, pattni_eco-evolutionary_2023, colnaghi_adaptations_2023}. However, many of these studies consider only a limited number of available environments, lack a generalized approach to defining the different environments and deal with game changes as a random process only, thus neglecting mechanisms leading to the selection of these environments \cite{hofbauer_evolution_2007,amaral_evolutionary_2016, venkateswaran_evolutionary_2018,stollmeier_unfair_2018, szolnoki_seasonal_2019, tilman_evolutionary_2020, roy_eco-evolutionary_2022, zeng_spatial_2022, han_novel_2023, chen_number_2023, feng_evolutionary_2023, kleshnina_effect_2023}. Thus, fundamental questions regarding how players' strategies and game environments mutually affect each other and co-evolve remain unanswered. In particular, the complex co-dependence between evolutionary strategy selection and evolutionary game selection dynamics itself, has not been studied previously.

In this work, we propose a new framework in which evolutionary dynamics acts both on strategies and game environments. This holistic perspective allows us to analyse how strategies and environments mutually co-evolve over time leading to the emergence of cooperative environments. Furthermore, we identify that the topological properties of social interaction networks play a crucial role in shaping individual behaviour. Thus, our framework encompasses a comprehensive exploration of both intrinsic (behavioural) and extrinsic (contextual) factors that contribute to the emergence of pro-social behaviour and social dilemmas.

The outline of the paper is the following. In Section~\ref{sec:model} we introduce the evolutionary model and explore its application to social dilemma games. We then look at the results of our model in a well-mixed population and networked structures in Section~\ref{sec:results}. Finally, we summarize our main findings and give some future directions in Section~\ref{sec:conclusion}.

\section{Model}\label{sec:model}

\begin{figure}
    \centering
    \includegraphics[width=0.85\columnwidth]{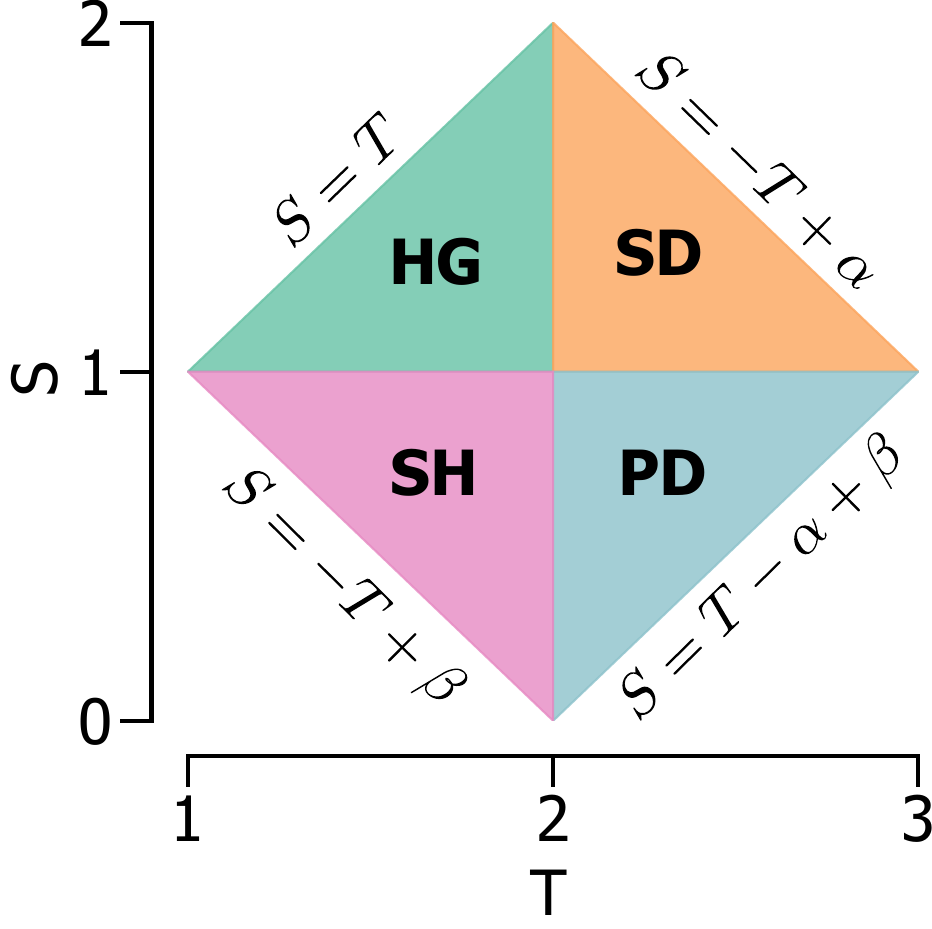}
    \caption{Games classification. Each player is assigned one pair of values ($T_i, S_i)$ from the game diamond. The equations for the boundaries of each game are derived from the inequalities between the payoffs. Here we set $\alpha = 4$, and $\beta=2$. }
    \label{fig:schematic}
\end{figure}

\subsection{Individual payoff matrices}

Let us consider two-player games where each agent can choose between two strategies. Traditionally, symmetric games, in the context of social dilemmas have been described by the following payoff matrix,
\begin{align}
   \kbordermatrix{
    & C & D \\
    C & R & S \\
    D & T & P  
  }
\end{align}
where, $C$ and $D$ represent the two available strategies: cooperation and defection, respectively. 

In particular, when facing another cooperator, a cooperator receives a payoff of $R$  (namely, ``reward''), while against a defector, the payoff is $S$ (referred to as ``sucker's payoff''). Conversely, a defector earns a payoff of $T$ (``temptation'') when facing a cooperator, and $P$  (``penalty'') when facing another defector.

The ordering of the entries of the payoff matrix defines four different types of games, namely:

\renewcommand{\jot}{10pt}
\begin{flalign}
\label{eq:games}
&\text{Prisoner's Dilemma (PD): } T>R>P>S \nonumber \\ 
&\text{Snowdrift (SD): } T>R>S>P \nonumber \\ \vspace{3em}
&\text{Stag-hunt (SH): } R>T>P>S  \\ \vspace{3em}
&\text{Harmony Game (HG): } R>T>S>P \nonumber
\end{flalign}
\renewcommand{\jot}{3pt}

In particular, we notice that for each game, either $T>R$ or $T<R$ and $S>P$ or $S<P$, leading to four distinct quadrants in the $T$-$S$ phase space.
In our framework, differently from the classical approach where just one game (i.e. a single payoff matrix) is taken into account, we consider a distribution of games. To do so we associate to each player $i$ a randomly generated game/payoff matrix. 
We note that the games can alternatively be represented in terms of `dilemma strengths' $D_g (=T-R)$ and $D_r (=P-S)$ instead of $T$ and $S$ \cite{tanimoto_relationship_2007, wang_universal_2015, ito_scaling_2018}. The two representations are equivalent under translational and rotational symmetry in the phase space.
If we define different games by generating the entries of the payoff matrix completely at random, the total payoff (i.e. the sum of the payoff matrix entries) for each game will, in general, be different. This introduces a bias, where the players associated with the games with a larger total payoff are favoured. Since in our model not only the strategies but also the games evolve, this bias will lead the evolutionary dynamics towards a trivial outcome where the whole population plays the game with the largest total payoff (as we verified numerically).
In our model, we remove this bias by considering games described by payoff matrices where the matrix entries (i.e. the payoffs) are drawn from symmetric distributions, and their sum (total payoff) is fixed.
This is achieved by using the following payoff matrix to define the game associated with player $i$:

\begin{align} \label{eq:payoff_matrix}
   \kbordermatrix{
    & C & D \\
    C & \alpha-T_i & S_i \\
    D & T_i & \beta-S_i  
  }
\end{align}
where $T_i \in [\beta/2, \alpha - \beta/2]$ and $S_i \in [\beta- \alpha/2, \alpha/2]$ are continuously distributed as shown in Fig.~\ref{fig:schematic}.

\begin{figure*}[!ht]
    \centering
    \includegraphics[width=2\columnwidth]{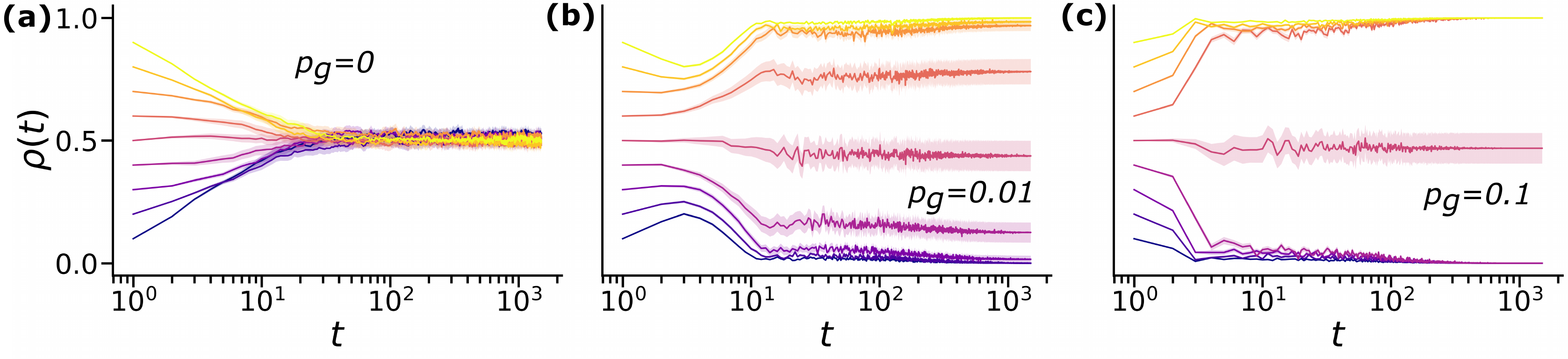}
    \caption{Fraction of cooperators $\rho(t)$ over time in a well-mixed population for \textbf{(a)} $p_g = 0$, \textbf{(b)} $p_g= 0.01$, and \textbf{(c)} $p_g= 0.1$.  The solid lines denote the averages while the shaded region is the standard error over 64 independent runs of the co-evolutionary game dynamics. Here we set $p_s=1$, $N=2500$, and $k=4$.}
    \label{fig:WM_timeseries}
\end{figure*}
In this way, the sum of the elements in each column of the payoff matrix (Eq.~\ref{eq:payoff_matrix}) is equal to $R+T=\alpha$ for the first column, and to $S+P = \beta$ for the second column. 
As a consequence, for fixed values of $\alpha$ and $\beta$ the total payoff (sum of all matrix entries) is a constant equal to $\alpha + \beta$.
By varying the values of $T_i$ and $S_i$ we can define all four types of games.
In particular, we can verify this by substituting the expressions $R=\alpha-T$ and $P = \beta-S$ into the payoff inequalities given by Eqs.~\ref{eq:games}. This allows us to derive the conditions for the different social dilemmas, while adhering to the constraint of a constant total payoff:
\begin{enumerate}
\item Prisoner's Dilemma (PD): $T>\alpha / 2$, $S<\beta/2$, and $S>T+\beta-\alpha$
\item Snowdrift (SD): $T>\alpha / 2$, $S>\beta/2$, and $S<\alpha-T$
\item Stag-hunt (SH): $T<\alpha / 2$, $S<\beta/2$, and $S>\beta-T$
\item Harmony Game (HG): $T<\alpha / 2$, $S>\beta/2$, and $S<T$
\end{enumerate}

We notice that $\alpha > \beta$ is a necessary condition for the existence of $T$ and $S$, solutions of the system of inequalities defining each game.
By representing these conditions in the $T$-$S$ space we obtain a ``games diamond"
as shown in Fig.~\ref{fig:schematic}.

\subsection{Co-evolutionary game dynamics}

We represent the state of player $i$ as $\Omega_i = (g_i,s_i)$. Each player $i$ within the population is distinctly identified by the game $g_i$ linked to it and its chosen strategy $s_i$, which can be either cooperation $C$ or defection $D$. As we saw in the previous section, given fixed values of $\alpha$ and $\beta$, the game $g_i$ is completely determined by the payoffs $T_i$ and $S_i$ and the payoff matrix Eq.~\ref{eq:payoff_matrix}. Henceforth, we will denote both the game and its corresponding payoff matrix by $g_i$.

The system evolves by agents replicating strategies $s_i$ and games $g_i$ from one another through an asynchronous update process \cite{schonfisch_synchronous_1999}. At each time step we randomly select a focal player $f$ and a model player $m$. 
We studied the evolutionary dynamics both for well-mixed populations, where each player can interact with all the other players, and for structured populations, where the players are represented as the nodes of a network and they interact if an edge connects them.
In particular, while in a well-mixed scenario the model player is randomly chosen from the entire population, in a structured population the model player is randomly drawn from the neighbours of the focal one.  

The focal (respectively, model) player collects total payoff $\pi_f$ (respectively, $\pi_m$) by playing with each of its opponents. In structured populations, the focal player and the model player interact with their $k_f$ and $k_m$ neighbours, respectively. In well-mixed populations, they engage with $ k $ randomly selected agents. Each player participates in two games against every opponent $ j $: one using their own payoff matrix $g_f$  ($g_m $ for the model), and another using the opponent's payoff matrix $g_j $. The focal player updates its strategy through a two-phase stochastic pairwise comparison process, through which he/she can adopt the strategy and \emph{the game} of the model player.

The state update consists of two steps. First, we determine whether the game, the strategy, or both will be updated: the game is selected for an update with probability $p_g$, and the strategy is selected with an independent probability of $p_s$. Note that the case $p_g=0$, $p_s=1$ recovers the usual evolutionary game dynamics where only the strategies evolve.
Given that environments usually change at a slower rate than the individual behaviours \cite{tilman_evolutionary_2020, mougi_eco-evolutionary_2023, cheng_evolution_2023}, we restrict ourselves to scenarios where $p_g \leq p_s$.
Then, the adoption of the model game and/or strategy occurs with a probability that increases with $\pi_m - \pi_f$, the payoff difference between the model and focal players. In particular, the probability of adoption is described by the so-called Fermi function \cite{schuster_replicator_1983, blume_statistical_1993, szabo_evolutionary_1998, santos_scale-free_2005}:
\begin{eqnarray}
    \Pi = \frac{1}{1 + \exp{(-w (\pi_m - \pi_f))}}\;,
\end{eqnarray} 
where $w$ governs the strength of selection. Note that the Fermi update rule is not deterministic and (especially for low values of $w$) the focal player can adopt games and strategies from the model player even when $\pi_m < \pi_f$.
Thus, at each step of the evolutionary dynamics, there are $4$ possible transitions from the original state of the focal player $\Omega_f(g_f,s_f)$, to a new state where both the strategy and the game can be those of the model player, each occurring with the following probabilities:
\begin{eqnarray}
 \Omega_f(g_f,s_f) \rightarrow 
    \begin{cases}
    \Omega_f(g_m,s_m)& \text{with } p_s\cdot p_g\cdot \Pi^2\\
    \Omega_f(g_m,s_f) & \text{with } p_g\cdot \Pi\left(1-\Pi p_s\right)\\
    \Omega_f(g_f,s_m) & \text{with } p_s\cdot \Pi\left(1-\Pi p_g\right)\\
    \Omega_f(g_f,s_f) & \text{with } \left(1-\Pi p_s\right)\left(1-\Pi p_g\right) \\
    \end{cases}
\end{eqnarray}

\begin{figure*}[t!]
    \centering
    \includegraphics[width=1.7\columnwidth]{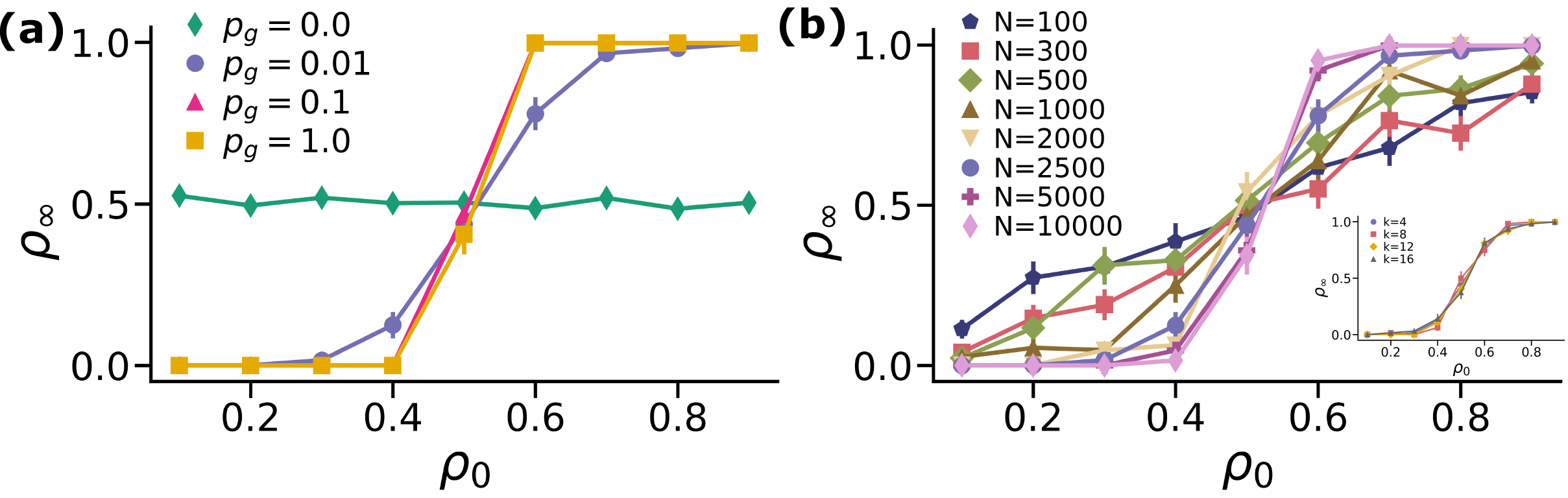}
    \caption{\textbf{(a)} Quasi-stationary state fraction of cooperators $\rho_{\infty}$ as a function of the initial cooperators' density $\rho_0$ for various values of $p_g$, the probability of selection of game update. The parameters are $p_s=1$, $N=2500$, and $k=4$. \textbf{(b)} Effect of size and degree (inset) on the quasi-stationary fraction of cooperators $\rho_{\infty}$ for $p_g=0.01$. All the results are for a well-mixed population and averaged over 64 runs.}
    \label{fig:newfig3}
\end{figure*}

We repeat the simulation step until the system reaches a quasi-stationary state in which we compute the relevant macroscopic order parameters. In stochastic processes featuring absorbing states, quasi-stationary methods are employed to identify the system's stable states \cite{QS_definition, de_oliveira_how_2005, zhou_evolutionary_2010}. If the system transitions into an absorbing state, the quasi-stationary methodology reverts it to a prior state, with a probability proportional to the time spent in that state. This approach yields a distribution (namely, quasi-stationary distribution) where the probability of each state is proportional to the time spent by the system in that state. Previous research has shown that the local maxima of the quasi-stationary distribution asymptotically approaches the stable fixed point inherent to the system's stochastic dynamics with growing system size \cite{faure_quasi-stationary_2014, sander_sampling_2016, civilini_evolutionary_2021}.

\section{Results}\label{sec:results}

We analyse the outcome of the co-evolutionary dynamics of games and strategies on different population structures. We first consider a well-mixed population of players, showing how the propensity to change the game, $p_g$, affects the system's steady state. In particular, we look at the fraction of cooperators $\rho$ in the quasi-stationary state as a function of time. Next, we consider the case of structured populations, where the interactions among players are represented as a network. In this way, we explore the impact on the model dynamics of fixed neighbourhoods (namely \emph{network reciprocity} effects), spatial homophily of games, and the correlations between the states of the agents and their placement in the network.

\subsection{Well-mixed population}
We consider well-mixed populations where we assign a game chosen uniformly at random from the game's diamond in Fig.~\ref{fig:schematic} to each player.  
We start our analysis of the model dynamics with the case $p_g = 0$ and $p_s=1$. In the rest of the paper we always choose $p_s=1$ unless stated otherwise. In Fig.~\ref{fig:WM_timeseries}(a) we observe that for $p_g = 0$, independently of the initial fraction of cooperators, the trajectories converge to a quasi-stationary state where the population cooperates roughly half of the time.
Thus, independently from the initial number of cooperators in the system, when $p_g = 0$ we always find players cooperating roughly half of the time. The error bars of the plot denote the standard error over 64 runs for each $\rho_0$.

In contrast to the scenario where $p_g=0$, a slight increase of $p_g$ breaks down the symmetry of the evolutionary outcome under changes in the initial configuration. For instance, even for a very small value of $p_g=0.01$ in Fig.~\ref{fig:WM_timeseries}(b), we observe a variety of quasi-stationary states in which the final density of cooperators strongly depends on the initial levels of pro-social behaviour (i.e., the initial fraction of cooperators). When $p_g$ is further increased the former effect becomes more pronounced and (see Fig.~\ref{fig:WM_timeseries}(c) for $p_g=0.1$) trajectories seem to bifurcate in two sets based on the number of initial cooperators. In particular, when the initial fraction of cooperators $\rho_0 = 0.5$, we observe that the cooperation level remains the same. However, if cooperators are in the minority, the system tends to eliminate cooperators while, when we start with a majority of cooperators, cooperation prevails in the long run. 

To have a better understanding of this effect we define $\rho_{\infty} = \lim_{t \rightarrow \infty} \langle \rho(t) \rangle$ as the time-averaged cooperators' density in the long time limit and explore the dependence of $\rho_{\infty}$ as a function of $\rho_0$ for various $p_g$ values. In Fig.~\ref{fig:newfig3}(a) we illustrate the symmetry-breaking phenomenon that occurs when $p_g>0$. In particular, as soon as $p_g>0$ we observe that a critical initial fraction of cooperators of $\rho_0$ $\approx 0.5$ is needed to have a majority of cooperators in the steady state of the system. Moreover, as $p_g$ increases the curves $\rho_{\infty}$ depend more and more non-linearly on $\rho_0$, resembling a step function when $p_g\rightarrow 1$. In this limit of strong game selection, the form of the curves $\rho_{\infty}$ pinpoint a reinforcement effect in which a slight bias towards cooperation (defection) in the initial configuration leads to a dynamical reinforcement of cooperators (defectors). 

Fig.~\eqref{fig:newfig3}(b) shows the dependence of system size $N$ on the cooperators' density in the quasi-stationary state, $\rho_{\infty}$ for $p_g=0.01$. We notice that the transition becomes sharper as the system approaches the thermodynamic limit $N\rightarrow\infty$. The finite size effects become significantly less relevant beyond $N \sim 1000$. Subsequently, we fix $N=2500$ for the rest of the manuscript. Fig.~\eqref{fig:newfig3}(b)(inset) shows the effect that the average number of interactions has on the degree of cooperation for $N=2500$ and $p_g=0.01$. We see that there is no effect on the quasi-stationary state density for different values of the average degree. Consequently, we fix $\langle k \rangle = 4$ for the rest of the manuscript unless stated otherwise.

\begin{figure}[!htp]
    \centering
    \includegraphics[width=0.9\columnwidth]{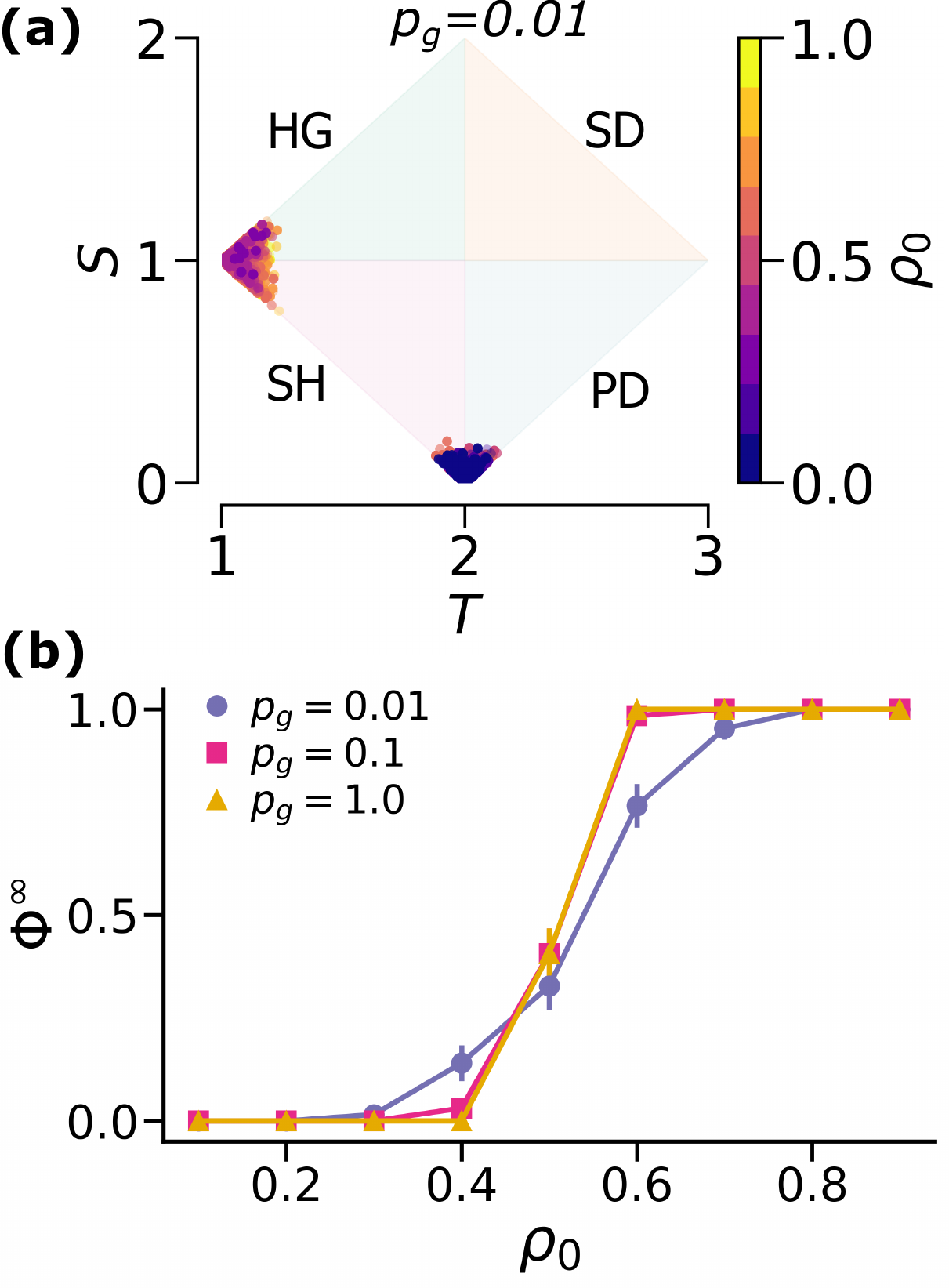}
    \caption{Game selection in well-mixed populations. \textbf{(a)} Surviving games in a well-mixed population for various values of initial cooperators. Violet points denote initial conditions with fewer cooperators than defectors, while yellow dots denote more initial cooperators. \textbf{(b)} Surviving cooperative games fraction $\Phi^{\infty}$ as a function of $\rho_0$ for the same values of $p_g$. The parameters are $p_s=1$, $N=2500$, and $k=4$.}
    \label{fig:WM_critmass}
\end{figure}

Now we analyse the effects of the co-evolutionary dynamics in game selection when $p_g>0$. In particular, in Fig.~\ref{fig:WM_critmass}(a), we show the distribution in the $T$-$S$ plane of the surviving games, where the colours denote the number of initial cooperators, $\rho_0$, of the corresponding realization. First, we observe that the quasi-stationary state consists primarily of players playing the SH, and also PD or HG depending on $\rho_0$. Moreover, it is interesting to notice that the surviving games are densely distributed near the corners of the phase space. In particular, we observe a transition of surviving games from PD-SH to HG-SH as a function of $\rho_0$. Fewer initial cooperators lead to games closer to PD, which has full defection as the Nash equilibrium. On the other hand, an initial majority of cooperators brings the system closer to HG, which has full cooperation as the Nash equilibrium. 

Finally, we analyse in depth the dependence of the selected games as a function of $p_g$ and $\rho_0$. The game $g_i$ played by each individual $i$ is defined by the values of $T_i$ and $S_i$ and can be classified as cooperative or non-cooperative. To this aim, we classify a game $g_i$ as cooperative if the associated Nash equilibrium (mixed or pure) has a majority of cooperators, otherwise $g_i$ is regarded as non-cooperative. In particular, a cooperative game holds when $S>T-1$. Figure \ref{fig:WM_critmass}(b) shows the fraction of cooperative games (environments) $\Phi^{\infty}$ among the surviving games as a function of $\rho_0$ for different values of $p_g>0$. It is interesting to see that the curves are quantitatively very similar to the trend of $\rho_{\infty}$ shown in Fig.~\ref{fig:newfig3}(a), implying a strong correlation between the selected games and strategies for all values of $p_g$.

In conclusion, the presence of co-evolutionary dynamics in well-mixed populations provides a way out for the survival of cooperation. However, this effect requires that $\rho_0 \approx 0.5$, which is an unrealistic critical mass of initial cooperators to achieve pro-social behaviour. In the next section, we show how the presence of structured populations can enhance pro-social behaviour at a lower cost (i.e. initial cost of cooperation).


\subsection{Structured populations}

Well-mixed populations are not the best representation of real-world systems, since individuals do not interact at random, but according to well-defined structural patterns. Networks constitute a natural framework for analysing these interaction paradigms. Here, we study the impact of different network topologies and features found in real-world systems on the model dynamics. 

\begin{figure*}[htp]
    \centering
    \includegraphics[width=1.8\columnwidth]{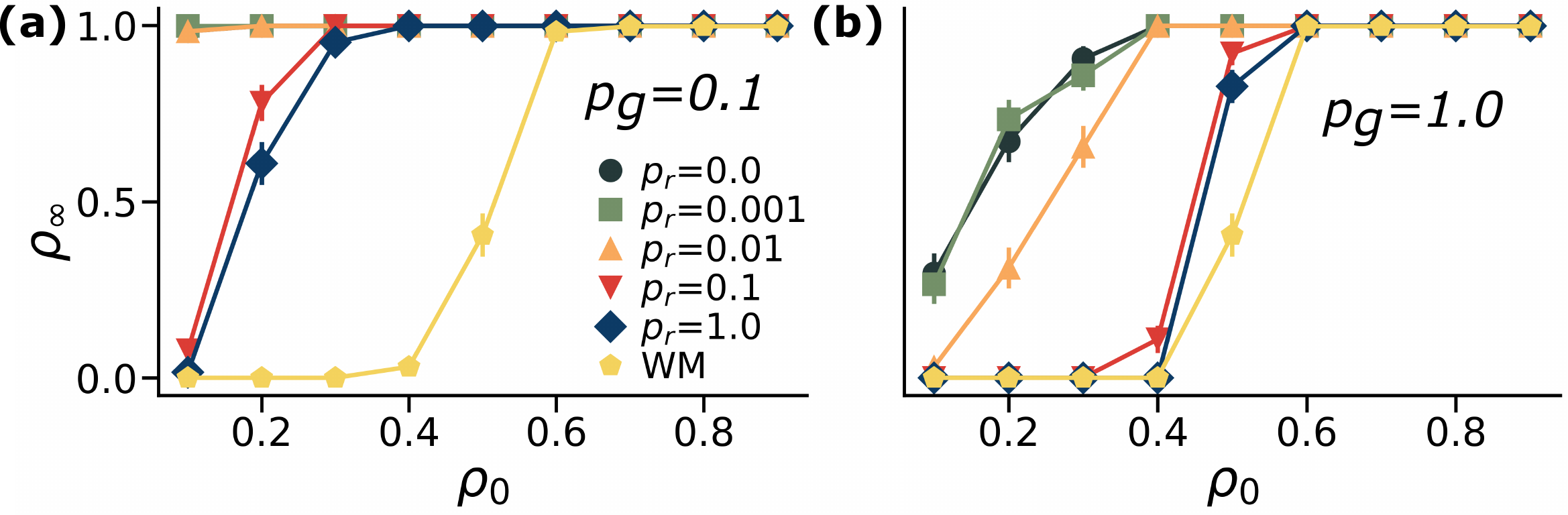}
    \caption{Fraction of cooperators in the quasi-stationary state $\rho_{\infty}$ as a function of $\rho_0$ for \textbf{(a)} $p_g=0.1$ and \textbf{(b)} $p_g=1$ for different rewiring values $p_r$ in a lattice along with well-mixed populations for comparison. Simulations are for populations of $N=2500$ individuals and $p_s=1$ averaged over 64 runs.}
    \label{fig:network}
\end{figure*}

As explained above, the use of well-mixed populations prevents repeated interactions between players since model players are chosen at random by each focal agent at each time step. Moreover, the opponents of the focal and model players are also drawn randomly from the population at each time step. However, when networks are used to represent the population structure, the neighbours of each agent are fixed. This implies that the set of possible model players for each focal player, and the set of possible opponents of each node, do not change in time. We will show that this can greatly impact the update of both strategies and games.
In the context of evolutionary game dynamics it is widely known that networked interactions provide a structural way to sustain cooperation through various mechanisms such as punishing those who defect, clustering the cooperators together, reinforcement of pro-social behaviour \cite{nowak_five_2006, santos_evolutionary_2006, gomez-gardenes_dynamical_2007, szabo_evolutionary_2007, amaral_evolutionary_2016, cheng_evolutionary_2023, fijalkow_games_2023, pires_network_2023}. This effect, also known as `network reciprocity', can lead to pro-social behaviour even in situations where, without network structure, cooperation can not be sustained.


Inspired by this phenomenon, we look at how a structured population can result in an enhancement of cooperative behaviour in the context of evolutionary game selection. Specifically, we are interested in how interacting with fixed neighbours and the structure of different network topologies change the quasi-stationary state compared to a well-mixed population. In the next part, we will explore how clusters of similar types of players (in our case having identical strategies and games) can emerge and how the targeted placement of cooperators on heterogeneous networks can amplify pro-social behaviour.

\subsubsection{The effect of network structures}

We start by examining co-evolutionary dynamics on 2D lattices, a topology that was first investigated in the context of evolutionary games by Nowak and May~\cite{nowak_evolutionary_1992}, as these structures can be easily represented as players on top of a surface, allowing us to easily visualize and investigate how spatial correlations affect the emergence of collective behaviour \cite{szolnoki_seasonal_2019, kumar_evolution_2020, flores_cooperation_2022, ichinose_how_2022, locodi_effects_2023}. Initially proposed by Watts and Strogatz~\cite{watts_collective_1998}, the link-rewiring mechanism systematically alters network structure going from a well-ordered periodic structure (lattice) to a disordered random structure displaying the so-called `small-world' phenomenon, where all agents in the system are at most few steps far apart from one another. By using a variation of the Watts-Strogatz model where we tune the probability of rewiring the edges $p_r$ in such a way that the degree of each node remains fixed, we change the local (clustering) as well as the global (shortest path length) structural properties, thus breaking the locally homogeneous patterns inherent to simple lattices

As we have shown previously, cooperators' density in the quasi-stationary state is highly correlated to the surviving game. Hence, from now on we only show the fraction of cooperators in the quasi-stationary state, $\rho_0$, since the trend for the fraction of surviving cooperative games is practically the same.
Figure~\ref{fig:network} shows the the fraction of cooperators in the steady state $\rho_{\infty}$ for various values of $p_r$ and two values of $p_g$. We also report the results for well-mixed populations for comparison.

For the smallest value of $p_g$ (Fig.~\ref{fig:network}(a)), well-mixed populations showcase a sigmoidal curve with the onset of cooperation around $\rho_0 \approx 0.4$. However, for rewired lattices, $\rho_{\infty}$ is consistently higher than that of well-mixed populations when we start from a minority of initial cooperators. In particular, we notice that for very small values of $p_r$ we always get a full cooperative state independent of the initial density of cooperators. Interestingly for intermediate values of $p_r$, we can get a majority of cooperators with around $20\%$ or fewer players starting as cooperators. However, the difference between random regular networks ($p_r=1$) and well-mixed populations is quite large, pointing to other possible mechanisms for the enhancement of pro-social behaviour beyond lattice rewiring.

The former picture changes dramatically when game selection is fully activated, $p_g=1$, as shown in Fig.~\ref{fig:network}(b). Differently from the previous case, when $p_g=1$ random regular networks ($p_r=1$) exhibit a very similar trend to well-mixed populations. On the other hand, rewired lattices (intermediate values of $p_r$) even though sustain cooperation better than random regular networks in this case, are far from their performance as boosters of cooperation as for $p_g=0.1$.

In summary, the former results indicate that networks boost pro-social behaviour, elevating the cooperation level in the population. Interestingly, smaller but non-zero values of $p_g$ increase the cooperative behaviour to a greater extent in locally ordered homogeneous structures (lattices) than in the case of large $p_g$ values. 
These results point to a balance between the propensity of changing game and the heterogeneity of interaction patterns.

\subsubsection{Clustering games on a 2D lattice}\label{sec:lattice}

In the next two subsections, we focus on analysing the structural organization of agents on the interaction backbone provided by the graph, to gain a deeper understanding of the role of the interaction structure in co-evolutionary game dynamics.

\begin{figure}[H]
    \centering
    \includegraphics[width=\columnwidth]{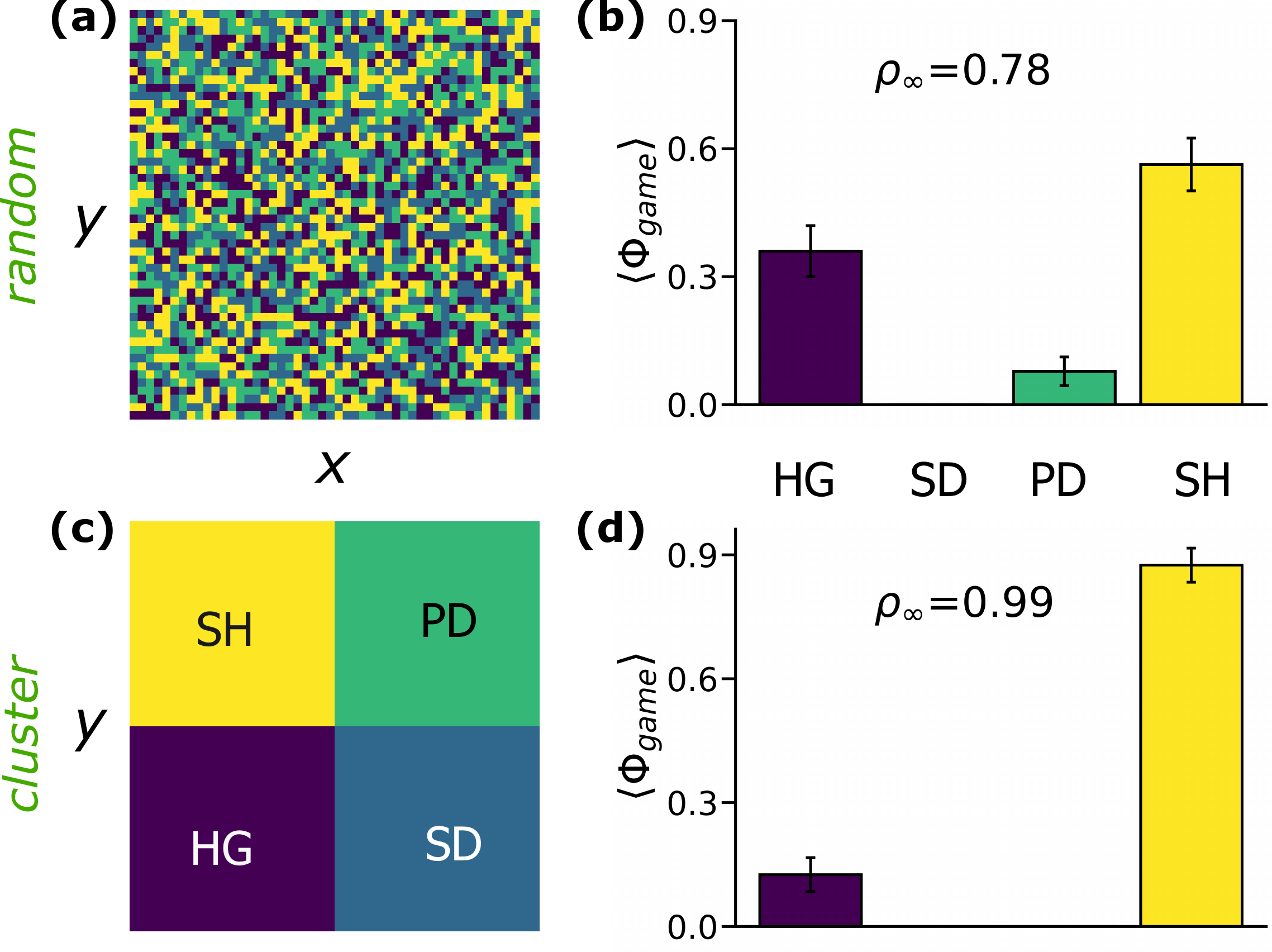}
    \caption{Surviving games on 2D square lattices for initial conditions with clustered or random games and $\rho_0=0.2$.  Initial snapshot of \textbf{(a)} randomly allocated games and \textbf{(c)} clustered games on a lattice of size $50 \times 50$. \textbf{(b), (d)} The fraction of players adopting a particular game type averaged over 64 runs in lattices. The text denotes the density of cooperators. Here, $p_g = p_s = 1$.}
    \label{fig:lattice}
\end{figure}

Figure~\ref{fig:lattice} depicts the initial snapshots of a 2D lattice when, initially, individual games are randomly placed, panel (a), or clustered, panel (c), such that the number of neighbours having the same game type as of a given node is significantly higher. Since we consider square lattices with periodic boundary conditions, all players are structurally equivalent and there are no boundary effects. In both cases, 20\% of the players are initially cooperators, located randomly on the lattice.

We quantify the quasi-stationary state distribution of games by defining $\langle \Phi_{game} \rangle$ as the average fraction of sites occupied by a given \textit{game}. We calculate the average over 64 independent runs and the error bars denote the standard error.

From panels Fig.~\ref{fig:lattice} (b) and (d) it becomes clear that SH games prevail the majority of times. When games are randomly placed, the Stag-hunt and the Harmony games dominate the final configuration of the lattice. In this particular scenario, Prisoner's Dilemma (PD) games also survive. When the games are initially clustered on the lattice, the small fraction of PD games disappear, while the proportion of the Stag-hunt games increases. The density of cooperators is further increased when games are clustered. Notably, the Snowdrift game always goes extinct, regardless of the initial configuration.

In conclusion, clustering games on a lattice seems to promote the prevalence of cooperative environments. Moreover, while cooperators' density is generally higher on a lattice compared to structures without spatial correlations, game clustering further amplifies the selection of cooperation-friendly environments. This behaviour is important since it further corroborates the effect of homophily in real-world systems \cite{mcpherson_birds_2001, karimi_homophily_2018}.

\subsubsection{Targeted placement of cooperators}\label{sec:sf}

Multiple empirical investigations have revealed the presence of heterogeneous and power-law degree distributions in social networks \cite{barabasi_emergence_1999, boccaletti_complex_2006}. A heterogeneous degree distribution implies the presence of \emph{hubs}, nodes with a particularly high degree respect to the average degree of the network.
Such nodes with high degrees play a central role in the dynamics taking place on the network by effectively disseminating information, opinions, and behaviours throughout the network \cite{santos_scale-free_2005, gomez-gardenes_dynamical_2007}. 
Here, we explore the effect of the strategic initial placement of a small fraction of cooperators (less than or equal to $0.005$ times the size of the population) on the evolution of pro-social behaviour. In particular, we consider configurations with cooperators initially placed on the hubs of the network. In addition, we investigate how changing the level of degree-heterogeneity of the network affects the co-evolutionary dynamics of the system for the placement strategy mentioned above.

\begin{figure}[H]
    \centering
    \includegraphics[width=\columnwidth]{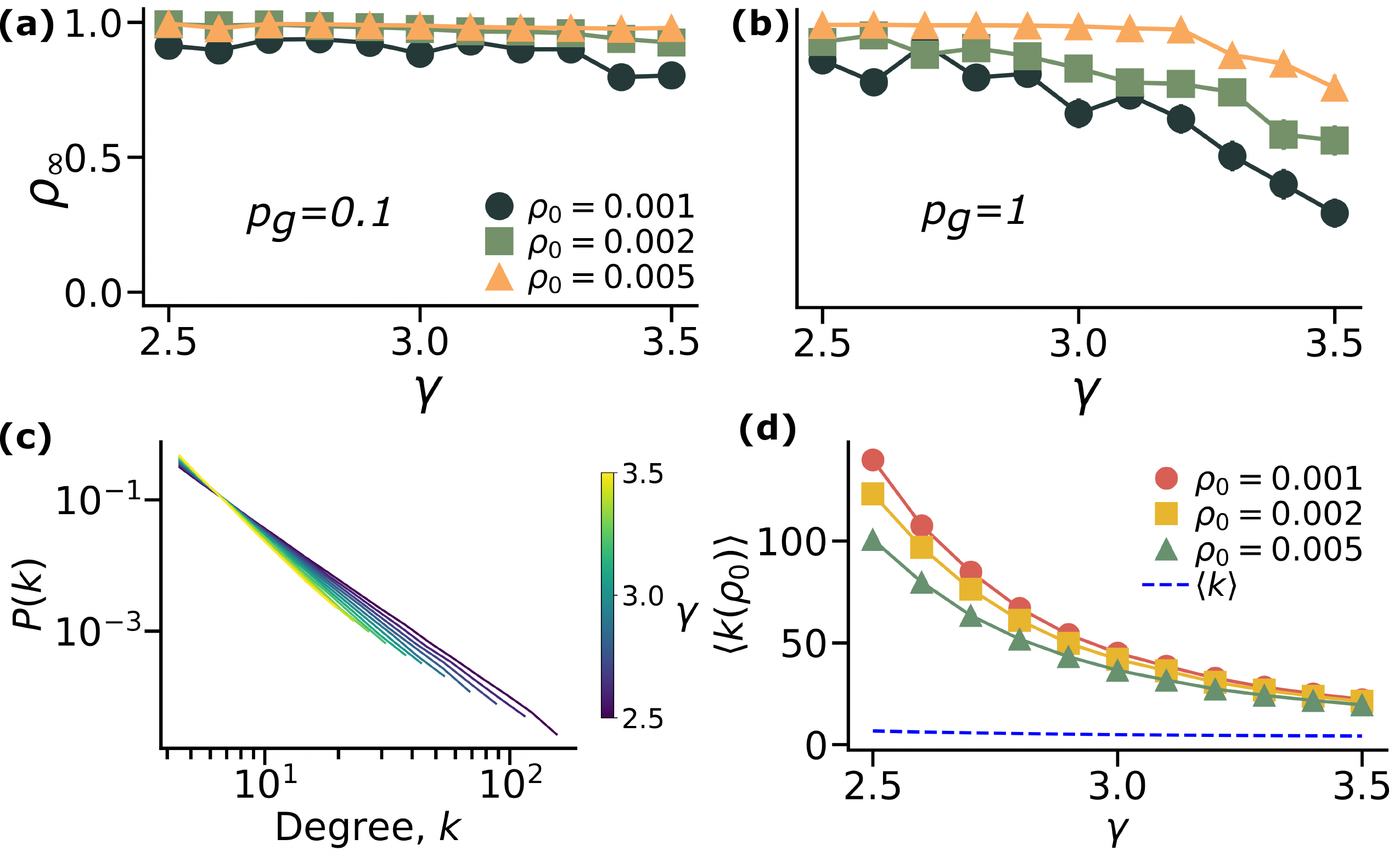}
    \caption{Cooperation levels in scale-free networks for strategic placements of cooperators. A small fraction of cooperators  (0.001-0.005 times the size of the population $N=2500$) is initially placed on the hubs for \textbf{(a)} $p_g=0.1$ and \textbf{(b)} $p_g=1.0$. \textbf{(c)} Degree distribution of scale free graphs for various values of $\gamma$ in a log-log scale. \textbf{(d)} The average degree of initial cooperators $\langle k (\rho_0) \rangle$ as a function of $\gamma$ for various values of $\rho_0$. The blue dashed line shows the average degree $\langle k \rangle$ of the networks.}
    \label{fig:scalefree}
\end{figure}

Figure~\ref{fig:scalefree} (a) $p_g=0.1$ and (b) $p_g=1$ shows the effect of strategically placing cooperators in power-law degree distributed networks as a function of their heterogeneity, captured by the value of the exponent $\gamma$ of the power-law.

For a small probability of game updating, \(p_g = 0.1\), placing cooperators on the hubs ensures a very high level of cooperation across all values of \(\gamma\) (Fig.~\ref{fig:scalefree}(a)). For example, starting with just three ($0.1\%$) cooperators among 2500 players leads to a final count of approximately 2000 cooperators ($\approx 80\%$). When game updating is always active, $p_g = 1$, and cooperators are placed on the hubs, cooperation levels for a similar initial cooperative mass decrease as the heterogeneity of the graph reduces (increasing \(\gamma\)), as illustrated in Fig.~\ref{fig:scalefree}(b).

To gain more insight into the mechanisms for the pro-social behaviour, we plot the degree distribution $P(k)$ for various values of $\gamma$ in Fig.~\eqref{fig:scalefree} (c). We see that $P(k)\sim k^{-\gamma}$, since the degree distribution follows a straight line in a log-log plot. Note that the slope of the distribution quantifies the heterogeneity and it decreases with increasing values of $\gamma$ \cite{boccaletti_complex_2006}. On the other hand, fig.~\eqref{fig:scalefree} (d) illustrates the average degree of the initial cooperators, $\langle k (\rho_0) \rangle$, as a function of $\gamma$. We observe that $\langle k (\rho_0 = 0.001) \rangle \sim 150$ for $\gamma = 2.5$, but it drops to $\sim 20$ for $\gamma =  3.5$.

To summarize, the collective behaviour of surviving games and strategies depends heavily on complex network features such as small-world behaviour, scale-free nature of the degree distributions, and the presence of strategy-degree correlations. In addition, increasing $p_g$ reinforces the effect of the degree heterogeneity in determining if cooperation is enhanced or diminished with respect to the strategic placement of initial cooperators on the hubs of the network.

\section{Conclusions}\label{sec:conclusion}

In the last decades, evolutionary game theory has provided valuable insights into understanding why agents choose cooperation despite personal incentives to defect. However, most existing studies focus on the evolution of strategies for specific, globally defined and static payoff matrices, disregarding changing environments and game conditions. Although some recent works have considered game heterogeneity through stochastic formulations, a comprehensive framework explaining the origin and emergence of these games and the dynamic relationship between games and strategies has remained elusive.

Our work contributes to bridging this gap by introducing a co-evolutionary framework where both strategies and games co-evolve and undergo evolutionary selection. We propose a simple model for game competition with various types of games ensuring an unbiased payoff distribution in all games. By adjusting the propensity to change the game ($p_g$) while maintaining a fixed probability of changing strategies ($p_s$), we discover fundamental changes in the system's evolution in well-mixed populations. In particular, even a small probability of switching games leads to a bifurcation of the quasi-stationary state, depending on the critical initial mass of cooperators. Additionally, we observe that the system tends to select more cooperative environments when there are enough initial cooperators, and the games and strategies influence each other, leading to a strong correlation between surviving games and strategies.

Beyond well-mixed scenarios, we find that structured populations enhance cooperation levels for small values of $p_g$. In particular, locally homogeneous graphs such as lattices lead to a state with a majority of cooperators even with a low initial mass of cooperators. However, this effect diminishes with the increase in the game selection propensity $p_g$ and the disruption of the regular lattice structure through a rewiring process, which creates shortcuts. In the specific case of 2D lattices, we also observed that clustering similar games promotes cooperation and leads to the survival of cooperative games.

In the case of scale-free networks we have found that the enhancement of cooperation decreases as the value of $p_g$ increases. The drastic difference in connection patterns of high-degree nodes compared to the rest of the nodes reinforces pro-social behaviour.

In summary, our findings shed light on the complex mechanisms shaping evolutionary processes and the interplay between strategic decision-making and mutating environments that define choices. Our work contributes to exploring the origins of social dilemmas prevalent in social settings. For the future, considering additional features such as community structure \cite{fotouhi_evolution_2019}, time-varying \cite{bhaumik_fixation_2023}, and higher-order interactions \cite{civilini_explosive_2024} may offer further insights into co-evolutionary processes of strategies and games in real-world systems \cite{traulsen_future_2023}. We hope that our work inspires more research on co-evolutionary dynamics as an avenue to tackle the puzzle of cooperation.

\section*{Acknowledgements}
A.C. and V.L. acknowledge support from 
the European Union -  NextGenerationEU, GRINS project (grant E63-C22-0021-20006).
F.B. acknowledges support from the Air Force Office of Scientific Research under award number FA8655-22-1-7025.
J.G.G. acknowledges financial support from the Departamento de Industria e Innovaweci\'on del Gobierno de Arag\'on y Fondo Social Europeo (FENOL group grant E36-23R) and from Ministerio de Ciencia e Innovaci\'on (grant PID2020-113582GB-I00).
The computational results presented have been achieved using the Vienna Scientific Cluster (VSC).

\bibliographystyle{apsrev4-2} 
\bibliography{references}

\end{document}